# Transition metal doped ZnS monolayer: The first principles insights

Rajneesh Chaurasiya and Ambesh Dixit

*Abstract*— Structural and electronic properties of pristine and transition metal doped ZnS monolayer are investigated within the framework of density functional theory. The pristine ZnS monolayer is showing direct band gap of about 2.8 eV. The investigated transition metal doping showed the transition from non-magnetic semiconductor to a magnetic system e.g. magnetic semiconductor for Co doped ZnS and half metal for Ni doped ZnS monolayers. The Co doped ZnS monolayer showed higher formation energy, confirming the strong bonding than that of Ni doped ZnS monolayer. The electron difference density shows the charge sharing between transition metal (Ni and Co) and S, confirming the covalent bond formation.

## 1   Introduction

Zn-VI compound semiconductors are considered as wide band gap semiconductors with hexagonal crystallographic structure. These materials are considered important for various applications including optoelectronics devices such as light-emitting diodes (LEDs), heterostructure photovoltaic devices, sensors and high power electronic devices [1, 2]. For example, ZnS thin film can be used as a buffer layer for solar cell application [3] because of its better adhesion with other compound semiconductors and high transmittance in the solar spectral region. The Zn-VI two dimensional structures are getting attention after experimental realization of graphene [4]. Along with computational investigations, there are experimental efforts as well simultaneously to realize and harness the potential two dimensional (2D) monolayer structures. ZnS nanosheets ~ 11 Å are synthesized in the wurtzite phase using hydrothermal method [5]. However, the graphene like ZnS hexagonal monolayer is not yet realized experimentally. The theoretical studies suggest that ZnS may exhibit monolayer in both planner and buckled structures [6]. The ZnS monolayer physical properties such as electronic, optical and mechanical etc [7,8] are investigated using the density functional theory, but only limited studies are available on doped ZnS monolayer [9]. Considering the same, present study aims to investigate the effect of TM such as Ni and Co doping on ZnS monolayer theoretically. The structural stability is analyzed through bond length, bond angle, formation energy and electron difference density. The electronic properties are considered using spin polarized generalized gradient approximation (S-GGA) to understand the impact of TM on electronic prop-erties of ZnS monolayer and Mulliken population analysis is used to understand the charge transfer in these systems.

## 2  Computational details

Spin polarized generalized gradient approximation (S-GGA) with Perdew-Burke-Ernzerhof (PBE) type parameterization is used for optimization and computing the physical properties of ZnS monolayer, as implemented in SIESTA code [10-12]. Norm conserving pseudopotentials of Zn, S, Ni and Co atoms are used in a fully separable form of Kleinman and Bylander in conjunction with double zeta double polarized basis sets. The force and stress convergence criterion considered are 0.001eV/Å and 0.01 eV/Å$^3$, respectively. A large plane wave mesh cut-off of 150 Ryd is used throughout the calculation. K point sampling of 10 x10 x1 is considered using Monkhorst-Pack scheme [13] along x, y and z directions, respectively for optimization and 20 x 20 x 1 used for the calculation of ground state properties. A 75 Ryd mesh cutoff is used for real space grid sampling.

## 3  Result and discussion

A 4 x 4 x 1 two dimensional hexagonal monolayer is considered for pristine and TM doped ZnS material with periodic boundary conditions in x-y directions and 15 Å vacuum is considered along z direction to avoid the interactions between the adjacent monolayers. The structural optimization of monolayer is carried out using force and stress criteria with 0.001 eV/Å and 0.01eV/Å$^3$ convergence limits, respectively. Lattice parameter and bond lengths 3.87Å and 2.23 Å, respectively for optimized pristine ZnS monolayer, which is showing graphene like planner structure, substantiating the sp$^2$ hybridization, shown in Fig.1. The results are in agreement with Zheng et al. work [14]. The lattice parameters are relatively insensitive to the transition metal doping, whereas a significant change in the bond length is observed. The formation energy is estimated as

$$E_f = E_{ZnS:TM} - E_{ZnS} + E_{Zn} - E_{TM}$$

Where $E_{ZnS:TM}$ and $E_{ZnS}$ are the total energies of transition metal doped ZnS and pristine ZnS monolayer, respectively. $E_{Zn}$ and $E_{TM}$ are the total energies of Zn atom and transition element, respectively. The computed formation energies are -3.22 and -2.65eV for Co and Ni doped ZnS monolayers. The relatively larger formation energy for Co doped ZnS monolayer substantiates the stronger bonding and stable configuration as compared to that of Ni doped ZnS monolayer. The computed electron difference densities are summarized with respective structures, confirming the electron sharing between cations and anions causing the formation of covalent bonds. The fractions of doped monolayer with pristine ZnS monolayer are shown in Fig 4 with respective electron difference densities for clarity. The difference densities clearly show the strong electron sharing between TM and neighboring S atoms. The electro-negativity of Ni and Co atom found to be ~ 1.88 for Co and ~1.91 and the relatively high electronegativity for Co is attributed to the observed enhanced charge sharing among nearby atoms as compared to that of the pristine and Ni doped ZnS.

Electronic properties of pristine ZnS monolayer are discussed through the band structure and density of states (DOSs) calculation. The respective band structures are plotted along the high symmetry Brillion zone (Γ-M-K-Γ) in the energy range of -5 to 5eV, setting Fermi energy at 0 eV. Conduction band minima and valance band maxima of pristine ZnS monolayer are located at Γ point and the observed energy difference ~ 2.8eV, supports the direct band semiconducting nature and consistent with Hamed et al. work [15]. We also computed the total and partial DOSs to understand the contribution of different orbitals in band structure. The valance band comprises of the S-p and Zn-d orbitals, which also hybridize with each other. The similar results are also observed from the electron difference density analysis, showing charge sharing between Zn and S atoms. The lower conduction band states are due to Zn-s orbitals, while the higher conduction band energy states are due to the Zn-p orbitals. The spin up and spin down states are confirming the nonmagnetic semiconductor behavior, in agreement with non-magnetic character of Zn and S atoms.

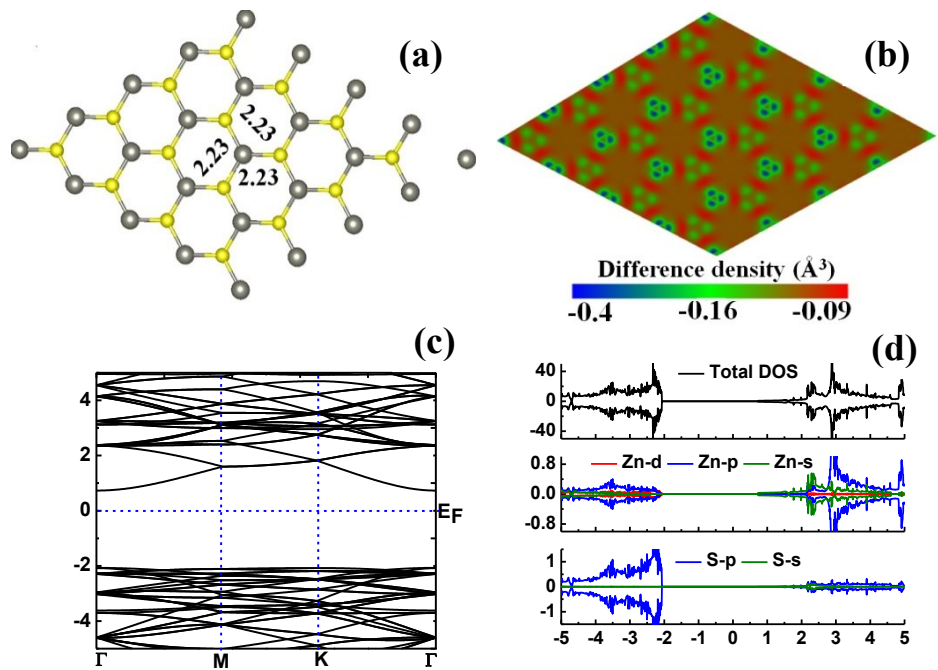

**Fig.1** (a) Optimized geometry (b) electron difference density (c) band structure and (d) density of states of ZnS monolayer

Fig.2 (a) reveals the optimized structure of Co doped ZnS monolayer, where one zinc atom is replaced by a Co atom. The observed Co–S bond length is 2.18Å, which is small than that of the pristine Zn-S bond length. The change in the bond length near the doped side is attributed to the different atomic radii of the dopant. The computed electron difference density, Fig.2 (b), is showing that Co atom acts as the charge accumulating center, which is shared with the nearby oxygen atom confirming the covalent behavior between Co-S. Fig.2 (c) reveals the band structure of Co doped ZnS monolayer, illustrating that spin down states causing an indirect band gap of 0.32eV while the spin up state has direct band gap of 2.35eV semiconductor behavior which confirms semiconducting behavior. The reduction and change in the band gap is observed due to the presence of excess bands in the CBM and VBM. The total and partial DOSs are computed and shown in Fig.2 (d), explaining that the contribution of spin down and spin up state. A large gap can be observed for spin up states, similar to that observed in electronic bands, Fig. 2(c), showing a semiconductor behavior for spin up electronic states. The VBM and CBM of spin down states are arising due to the presence of Co-d orbitals, while the spin up states for CBM and VBM are dominated due to the presence of the Zn-s and S-s orbitals, respectively. Co-d and S-s orbital contribute near the Fermi energy of valance band, substantiating the hybridization between them. Asymmetric DOSs confirmed the magnetic nature of Co doped ZnS monolayer with the observed total magnetic moment ~ 3$\mu_B$. The local magnetic moment of Co atom is ~ 2.42$\mu_B$ and rest is attributed to S atoms around Co atom, consistent with Juan et al. work on TM doped ZnO monolayer [16].The above studied electronic and magnetic properties confirm the ferromagnetic semiconductor properties, low spin states are contributing to the observed magnetic moment.

Fig.3 (a) shows the optimized structure of Ni doped ZnS monolayer and the relaxed structure of Ni doped ZnS monolayer showed the change in bond length around the dopant site as a result, the bond length of Ni-S is ~ 2.18Å which is smaller than pristine Zn-S bond length. Further, electron difference density is plotted in Fig.3 (b), showing that Ni atom is behaving similar to that of Co atom in ZnS monolayer, as discussed above. The sharing of charge between Ni and nearby oxygen atoms is visible form these electron difference density plots. The band structure is plotted in Fig. 3(c) for Ni doped ZnS monolayer, showing that metallic behavior for spin down state and insulating (or semiconducting) nature for the spin up state with a 2.41 eV direct band gap. The computed total and partial DOS are plotted in Fig.3 (d), showing the contribution of spin down and spin up state in bands formation.

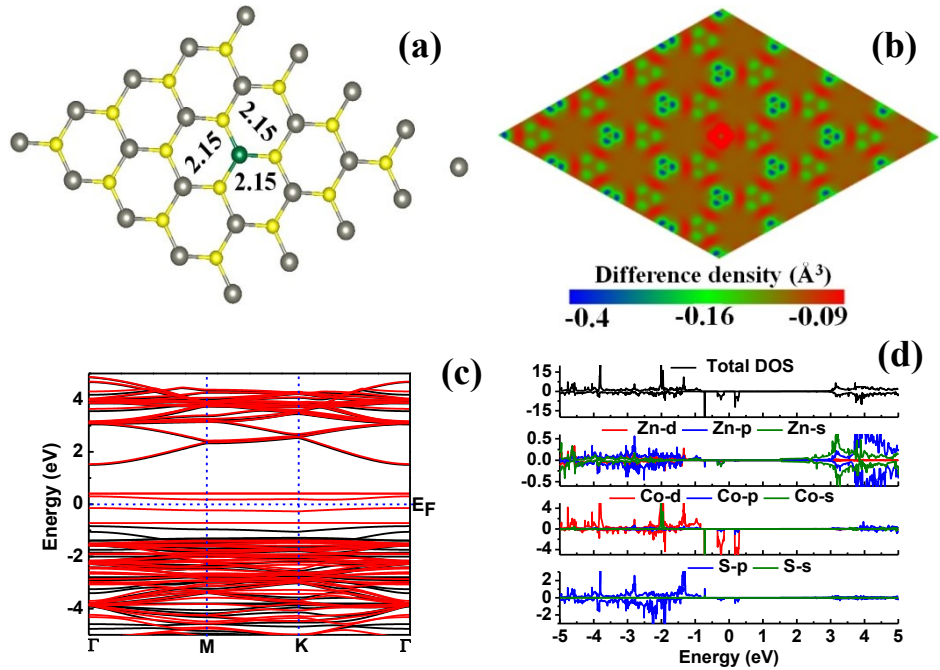

**Fig.2** (a) Optimized geometry (b) electron difference density (c) band structure and (d) density of states of Co doped ZnS monolayer

These observations suggest the metallic and semiconducting behavior for spin down and spin up states, respectively. The VBM and CBM of spin down states are due to presence of Ni-d and Zn-s orbitals, whereas spin up states in VBM are dominated by Ni-d and S-p, while the CBM is dominated by Zn-s orbitals. Ni-d and S-p orbitals contribute near the Fermi energy valance band, confirming the strong hybridization. The observed asymmetry in the spin down and spin up states, confirmed the magnetic behavior of Ni doped ZnS monolayer. The total magnetic moment is $2\mu_B$ which is the sum of Ni local magnetic moment ~ $1.42\mu_B$ and S atomic contribution around the Ni atom, consistent with Juan et al. work on TM doped ZnO monolayer [16]. The above studied electronic and magnetic properties confirm the ferromagnetic half-metal properties for Ni doped ZnS monolayer.

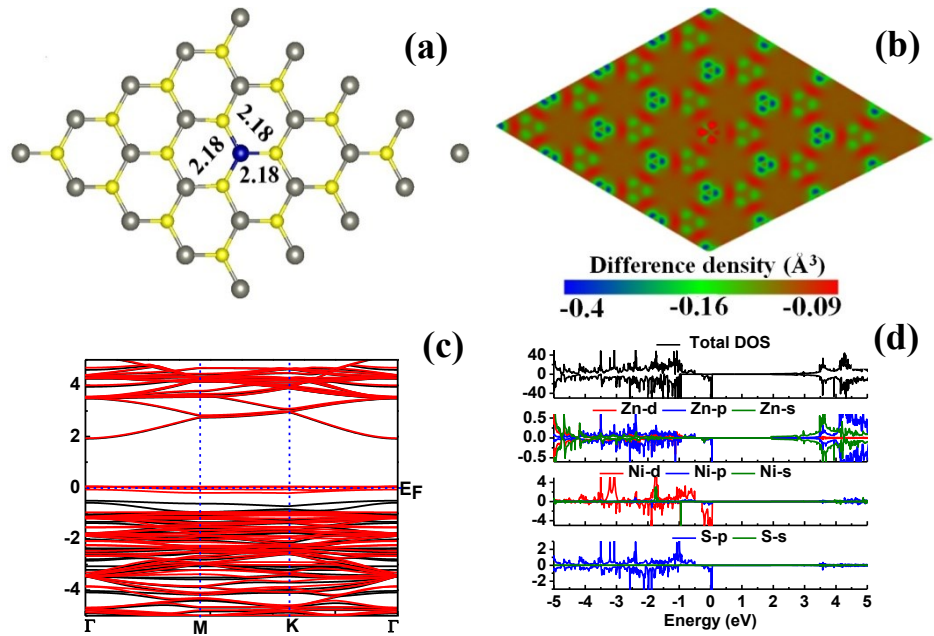

**Fig.3** (a) Optimized geometry (b) electron difference density (c) band structure and (d) density of states of Ni doped ZnS monolayer

Further, electron difference densities near doped transition metals in ZnS monolayers are summarized in Fig 4 together with pristine ZnS monolayer. The relative differences in ZnS monolayer suggests that localized distribution in comparison to that of strong sharing in case of transition metal doped ZnS monolayers. The sharing fraction is the maximum in case of Co doped ZnS monolayer, as shown by broadened electron difference density near Co atom. The situation is nearly similar for Ni doped ZnS monolayer, however there are subtle differences, causing the distribution of 3d Ni electrons near the Fermi energy, making it half metallic, as observed in PDOS, Fig. 3(d). Thus, Ni doped ZnS monolayer may provide 100% spin polarized electrons, making it suitable as spintronic material.

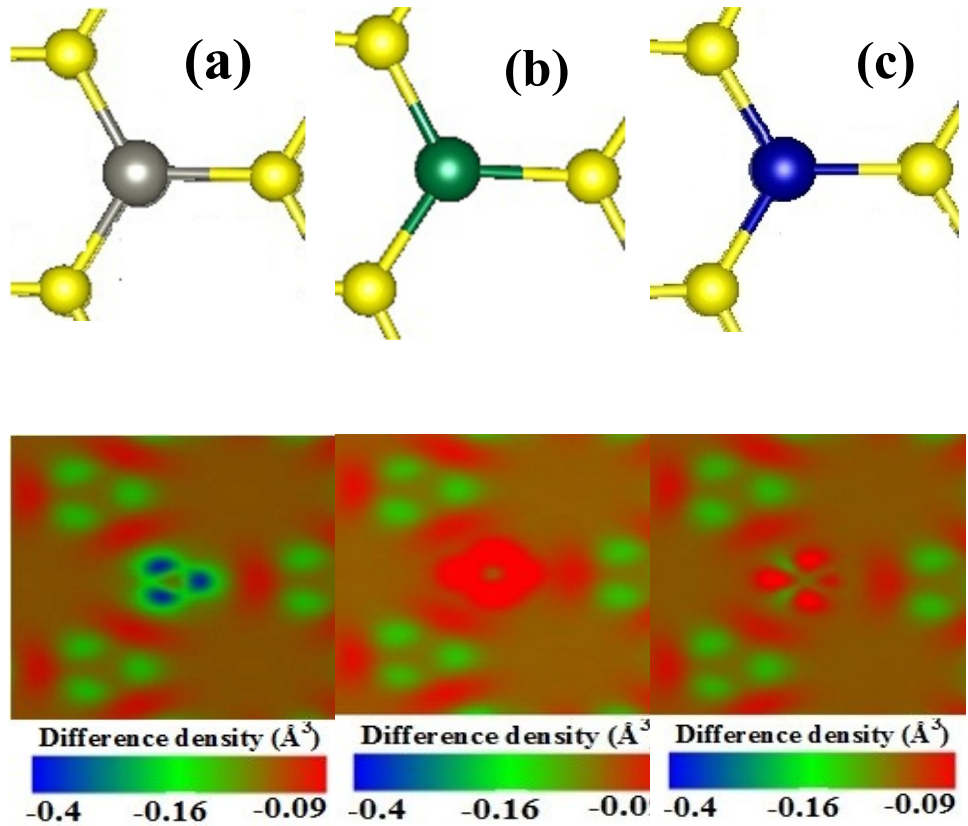

**Fig.4** (a) Pristine, (b) Co doped, and (c) Ni doped ZnS monolayer fraction near dopant site and the respective localized electron difference density plots.

## Conclusion

Pristine ZnS monolayer showed the charge sharing between the Zn and S atom without any effective magnetic moment. The doping of Co gives rise to a magnetic state, whereas doping of Ni showed a transition from non-magnetic to half-metallic state. Further, doping has showed the reduction in bond lengths near dopant site, with respect to that of the pristine ZnS. These studies provide an insight on structural and electronic properties of the pristine and doped ZnS monolayers. The nickel doped ZnS monolayer showed the half metallic behavior in contrast to magnetic semiconductor behavior for cobalt doped ZnS monolayer. These transition metal doped ZnS monolayers may be useful in realizing future spin electronic devices.